\documentclass[pra,aps,twocolumn,superscriptaddress,nofootinbib,nopacs]{revtex4} 

\usepackage{amsmath}  \usepackage{amssymb}  \usepackage{amsfonts}  \usepackage{bm}  \usepackage{bbm}  \usepackage{braket}  \usepackage{color}  \usepackage{comment}  \usepackage{dcolumn}  \usepackage{epsfig}
\usepackage[T1]{fontenc}  \usepackage{gensymb}  \usepackage{graphicx}  \usepackage[colorlinks,linkcolor=blue,citecolor=blue,urlcolor=blue,hyperindex,driverfallback=dvipdfm]{hyperref}  \usepackage{indentfirst}  \usepackage{lmodern}  \usepackage{mathrsfs}  \usepackage{mathtools}  \usepackage{psfrag}  \usepackage{pst-all}  \usepackage{soul}  \usepackage{xcolor}

\usepackage{float} 

\def\ii{{\rm i}}  \def\ee{{\rm e}}

  \def\ab{{\bf a}}    \def\bb{{\bf b}}  \def\Eb{{\bf E}}    \def\Fb{{\bf F}}  \def\fb{{\bf f}}  \def\Gb{{\bf G}}  \def\gb{{\bf g}}  \def\Hb{{\bf H}}          \def\kb{{\bf k}}      \def\Qb{{\bf Q}}    \def\Rb{{\bf R}}  \def\rb{{\bf r}}        \def\vb{{\bf v}} 
\def\xx{\hat{\bf x}}  \def\yy{\hat{\bf y}}  \def\zz{\hat{\bf z}}  \def\nn{\hat{\bf n}}  \def\rr{\hat{\bf r}}  \def\nn{\hat{\bf n}}    \def\RR{\hat{\bf R}}  
  \def\kparb{{\bf k}_\parallel} 
      
\def\mb{{\bf m}}

\begin{document} 
\def\bibsection{\section*{\refname}} 

\title{Coherent Smith-Purcell $\gamma$-Ray Emission
}


\author{Kamran~Akbari}
\affiliation{ICFO-Institut de Ciencies Fotoniques, The Barcelona Institute of Science and Technology, 08860 Castelldefels (Barcelona), Spain}
\author{Simone~Gargiulo}
\affiliation{Laboratory for Ultrafast Microscopy and Electron Scattering (LUMES), Institute of Physics, \'Ecole Polytechnique F\'ed\'erale de Lausanne (EPFL), Lausanne CH-1015, Switzerland}
\author{Fabrizio~Carbone}
\affiliation{Laboratory for Ultrafast Microscopy and Electron Scattering (LUMES), Institute of Physics, \'Ecole Polytechnique F\'ed\'erale de Lausanne (EPFL), Lausanne CH-1015, Switzerland}
\author{F.~Javier~Garc\'{\i}a~de~Abajo}
\email[Corresponding Author:~]{javier.garciadeabajo@nanophotonics.es}
\affiliation{ICFO-Institut de Ciencies Fotoniques, The Barcelona Institute of Science and Technology, 08860 Castelldefels (Barcelona), Spain}
\affiliation{ICREA-Instituci\'o Catalana de Recerca i Estudis Avan\c{c}ats, Passeig Llu\'{\i}s Companys 23, 08010 Barcelona, Spain}



\begin{abstract}
We investigate the Smith-Purcell emission produced by electron- or ion-beam-driven coherent excitation of nuclei arranged in periodic crystal lattices. The excitation and subsequent radiative decay of the nuclei can leave the target in the initial ground state after $\gamma$-ray emission, thus generating a coherent superposition of the far-field photon amplitude emanating from different nuclei that results in sharp angular patterns at spectrally narrow nuclear transition energies. We focus on Fe-57 as an example of two-level nuclear lossy system giving rise to Smith-Purcell emission at 14.4\,keV with a characteristic delay of 142\,ns relative to the excitation time. These properties enable a clean separation from faster and spectrally broader emission mechanisms, such as bremsstrahlung. Besides its fundamental interest, our study holds potential for the design of high-energy, narrow-band, highly-directive photon sources, as well as a means to store energy in the form of nuclear excitations.
\end{abstract}

\maketitle 
\date{\today} 

\section{Introduction}

The interaction of electron beams with periodic gratings enables the generation of coherent radiation over a wide spectral range in synchrotrons and free-electron lasers \cite{J99,SSY1}. Indeed, far-field radiative components are produced in general upon scattering of the evanescent electromagnetic field that accompanies electrons in motion when they cross or pass near a material structure, giving rise to cathodoluminescence emission \cite{paper149}. If the structure is periodically patterned with spatial period $d$ along the electron velocity vector $\vb$, Smith-Purcell (SP) radiation \cite{SP1953} is emitted over a broad range of wavelengths $\lambda$ and angles $\theta_n$ relative to $\vb$ satisfying the condition of far-field constructive interference
\begin{align}
c/v-\cos\theta_n=n\lambda/d, \label{SP}
\end{align}
where the integer $n$ labels different coherence orders. The SP effect has been extensively studied using macroscopic structures such as gratings and particle arrays \cite{V1973a,M92,I00,paper027,YIH02,K05,OO05,BDP08,RSR17,RKT19}, as well as atomic layers in stacked van der Waals materials \cite{paper356}. In addition, electrons act in unison if they are bunched within small beam regions compared to the emission wavelength in the moving frame, giving rise to superradiant light generation ranging from the terahertz \cite{UGK98} to the x-ray \cite{SR1989,AB04} spectral domains. Incidentally, the inverse process (electron energy gain near an illuminated grating) has also been demonstrated \cite{MPN1987} and explored as a route to realise laser-based table-top particle accelerators \cite{KKR95}.

Resonances in the periodic structure can enhance the emission, as demonstrated using plasmonic gratings \cite{paper252}. Likewise, atoms in a crystal structure can be resonantly excited by each passing electron, so that subsequent radiative decay to their initial ground states leaves the target unchanged and thus gives rise to a coherent superposition of amplitudes contributed by different atoms in the far field. The resulting wavelengths and directions of light emission are equally described by Eq.~\eqref{SP}.

Nuclear transitions can also be excited by electromagnetic interaction with free electrons \cite{WA1979} and ions \cite{ABH1956}, and an analogous SP effect responding to the same kinematic condition as with electronic transitions is expected to take place. Supporting this possibility, coherent photon scattering from arrays of nuclei is inherent in the M\"ossbauer effect \cite{VF98}, which has prompted further studies of nuclear-based coherent optical response \cite{HT99,S99}. Given the large variety of nuclear excitation energies extending from $\sim10\,$eV to megaelectronvolts, as well as the lifetimes of such excitations ranging from subnanoseconds to many years \cite{IAEA}, electron- or ion-driven excitation of nuclei in a solid crystal offers a unique platform for exploring new physics with potential application in extreme light sources. A particularly fascinating opportunity is open by the coherent superposition of the emission from different nuclei when their radiative decay occurs with a long characteristic delay time after the passage of the charged projectiles, so that the angular distribution of the generated $\gamma$ rays is controlled by the direction and velocity of the exciting beam.

In this work, we investigate coherent $\gamma$-ray emission associated with the nuclear excitations produced when relativistic electrons or ions traverse a periodic crystal target. Specifically, we focus on nuclear excitations in Fe-57 as an example of two-level lossy system in which the emission probability is relatively intense. The resulting SP emission takes place over a temporal scale dictated by the lifetime of these excitations (142\,ns), and therefore, it can be neatly separated from faster and more intense coherent emission mechanisms, such as bremsstrahlung (BR) \cite{SU1985}, which occurs during the projectile crossing of the target (e.g., within $<1\,$fs for a 100\,nm-thick crystal). Besides its fundamental interest, our work emphasizes the potential of nuclear SP emission for application in extreme high-energy photon sources, capable of producing strongly directional angular patterns over the relatively long times associated with the slow decay of the selected nuclear excitations.

\begin{figure}[htpb]
\centering
\includegraphics[width=0.45\textwidth]{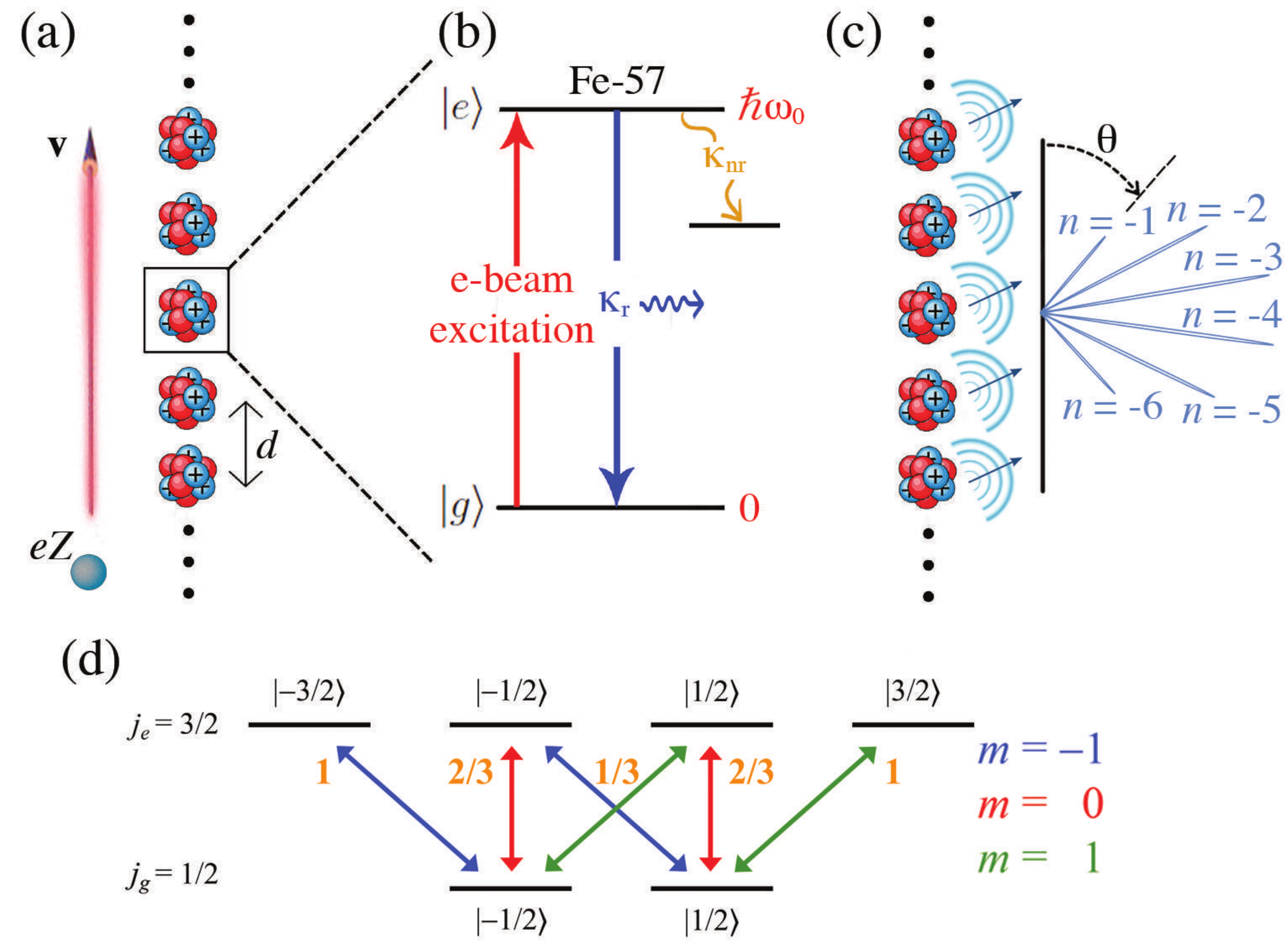}
\caption{\textbf{Smith-Purcell (SP) $\gamma$-ray emission.} (a) We consider an electron or ion of charge $eZ$ moving close and parallel to a linear periodic array of nuclei. (b) The moving charge can excite the nuclei, giving rise to radiative and nonradiative decay, here illustrated for Fe-57 (excitation energy $\hbar\omega_0=14.4\,$keV, total and radiative lifetimes $\kappa^{-1}=142$\,ns and $\kappa_r^{-1}=2.03\,\mu$s). (c) The resulting radiative emission is coherent if the nuclei return to their original ground state, thus producing $\gamma$ rays along angles $\theta$ relative to the projectile velocity $\vb$ satisfying the Smith-Purcell condition of far-field constructive interference [Eq.\ (\ref{SP})] at integer orders $n$ (see profile calculated for $v=0.94\,c$ and 10 nuclei in a linear array of period $d=2.86\,${\AA}). (d) Radiative transitions between different excited- and ground-state sublevels in Fe-57 have relative strengths (see labels) that depend on the azimuthal angular number $m$ of the emitted photon (color-matched arrows).}
\label{Fig1}
\end{figure}

\section{Results and discussion}

A fast electron or ion passing parallel to a periodic array of nuclei [Fig.~\ref{Fig1}(a)], such as those arranged in a solid crystal, induces nuclear excitations that can subsequently decay via radiative and nonradiative processes [Fig.~\ref{Fig1}(b)]. If the nuclei return to their initial ground states, the coherent superposition of the associated radiative emission amplitudes gives rise to a sharp angular pattern, as prescribed by Eq.~\eqref{SP} [Fig.~\ref{Fig1}(c)]. Here, we consider the $\hbar\omega_0=14.4\,$keV nuclear excitation in Fe-57 (see Appendix for an analysis of the 43.8\,keV excitation in Dy-161), in which both ground and excited states are degenerate as illustrated in Fig.~\ref{Fig1}(d), where sublevels are labeled by their respective angular momentum numbers $\mu_g$ and $\mu_e$ with $|\mu_g|\le j_g=1/2$ and $|\mu_e|\le j_e=3/2$. Coherent emission is produced if the excited nucleus decays back to its original ground state, whereas incoherent photons are generated otherwise. Transitions are dominated by the magnetic dipole channel, assisted by the absorption or emission of photons with an angular momentum number $m=\mu_e-\mu_g$. Averaging over the two possible $\mu_g$ sublevels, the combined process of excitation and subsequent de-excitation is coherent in a fraction $f=2/3$ of the interaction events. In addition, internal conversion, whereby the excited state donates its energy to an electron in the system, produces nonradiative decay with a probability that is $\alpha_{\rm IC}=8.544$ times higher than for radiative decay \cite{KBT08,BrIcc}. This leads to a radiative lifetime $\kappa_r^{-1}=\kappa^{-1}(1+\alpha_{\rm IC})/f=2.03\,\mu$s, where $\kappa^{-1}=142\,$ns is the measured lifetime of the excited state \cite{IAEA}.

Combining these elements, we can describe the SP emission under consideration by assigning a frequency-space induced magnetic moment $\mb_j(\omega)=\alpha_M(\omega)\Hb^{\rm ext}(\rb_j,\omega)$ to each nucleus $j$ in the structure (see details in the Appendix), where
\begin{align}
\alpha_M(\omega)\approx\frac{3}{4k^3}\frac{\kappa_r}{\omega_0-\omega-\ii\kappa/2} \label{alphaM}
\end{align}
is an isotropic $\mu_g$-averaged magnetic polarizability, $\Hb^{\rm ext}(\rb,\omega)=(2eZ\omega/vc\gamma)K_1(\omega b/v\gamma)\ee^{\ii\omega z/v}\hat{\boldsymbol{\varphi}}$ is the magnetic field generated at the nuclear position $\rb_j$ by a passing projectile of charge $eZ$ (e.g., $Z=-1$ for electrons) and velocity $v$ \cite{paper149}, $k=\omega/c$ is the light wave vector, and $\gamma=1/\sqrt{1-v^2/c^2}$ is the relativistic Lorentz factor. Considering the electric far-field $\Eb^{\rm ind}(\rb,\omega)=k^2\,\mb_j(\omega)\times\rr\,\ee^{\ii k|\rb-\rb_j|}/|\rb-\rb_j|$ induced by each magnetic dipole $\mb_j(\omega)$, and summing over the contributions from all nuclei, we find a probability of coherently emitting one $\gamma$ photon to be given by (see Appendix) $\Gamma^{\rm coh}=\int d^2\Omega\;\Gamma^{\rm coh}(\Omega)$, where
\begin{align}
\Gamma^{\rm coh}(\Omega)=\frac{9Z^2\alpha}{8\pi(v/c)^2\gamma^2}\;\frac{\kappa_r^2}{\omega_0\kappa}\;\left|\rr\times\gb(\Omega)\right|^2
\nonumber
\end{align}
is the angle-resolved probability, $\Omega=(\theta,\varphi)$ indicates the direction of $\rr$, and $\alpha\approx1/137$ is the fine structure constant. Here, we have introduced the dimensionless far-field amplitude
\begin{align}
\gb(\Omega)=\sum_j K_1\left(\frac{\omega_0|\Rb_j-\Rb_p|}{v\gamma}\right)\,\ee^{\ii\omega_0z_j/v}\,\ee^{-\ii\kb_0\cdot\rb_j}\,\hat{\boldsymbol{\varphi}}_{jp},
\label{gO}
\end{align}
where $\kb_0=(\omega_0/c)\,\rr$, $\Rb_j$ denotes the components of $\rb_j$ in a plane perpendicular to the velocity vector $\vb$, $\Rb_p$ defines the point of crossing of the probe in that plane, and $\hat{\boldsymbol{\varphi}}_{jp}$ is an in-plane unit vector perpendicular to $\Rb_j-\Rb_p$.

\begin{figure}[htpb]
\centering
\includegraphics[width=0.45\textwidth]{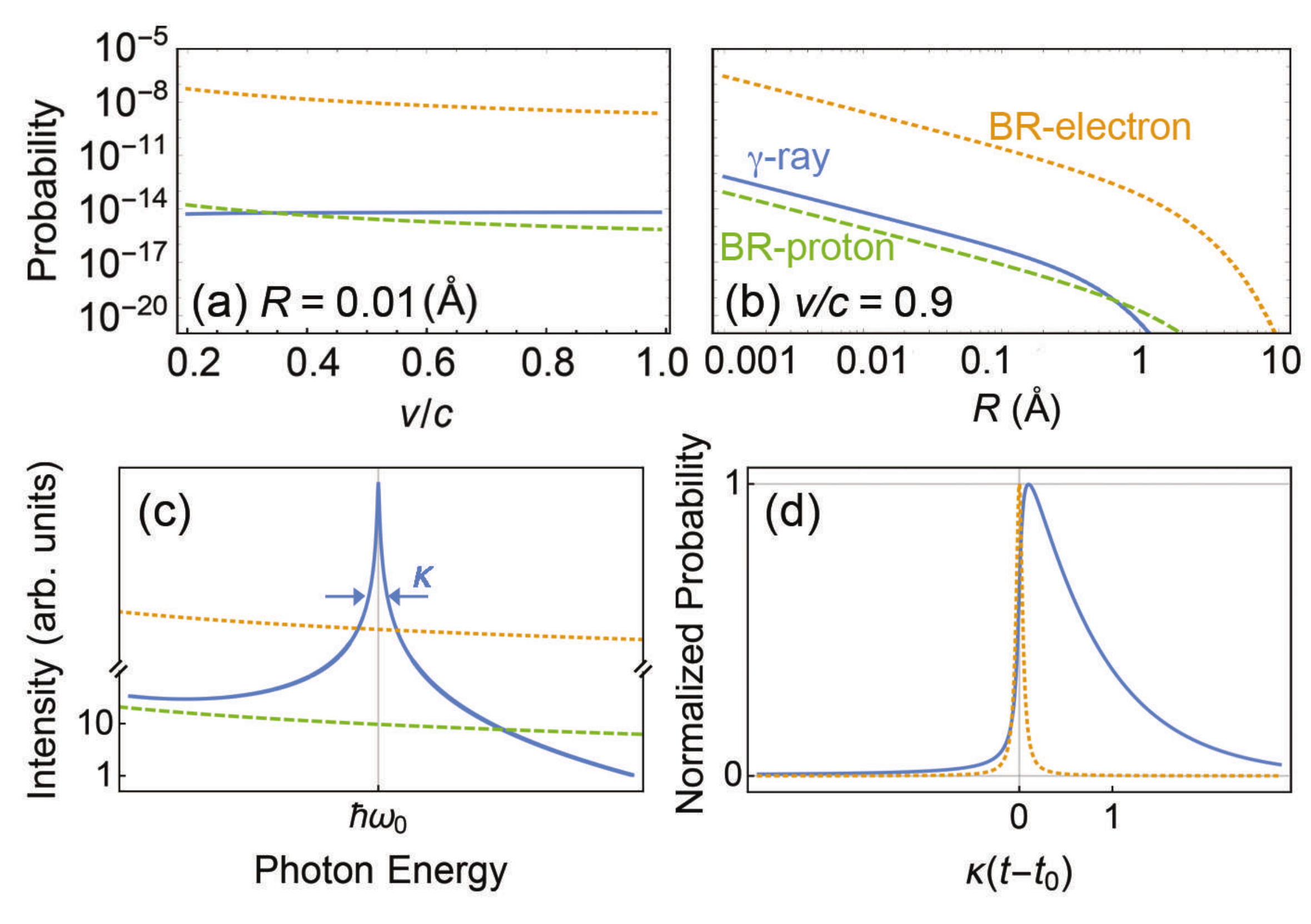}
\caption{\textbf{Photon emission from an individual nucleus.} (a,b) Probability of $\gamma$-ray emission (solid curves) compared with BR emission (broken curves, integrated over 1\,eV around $\hbar\omega_0$) for electrons and protons passing at a distance $R$ from an individual Fe-57 nucleus as a function of either probe velocity $v$ for $R=0.01\,${\AA} (a) or distance for $v/c=0.9$ (b). (c) Spectral profile of the emission associated with $\gamma$-ray and BR processes. (d) Temporal dependence of the normalized $\gamma$-ray and BR emission probabilities.}
\label{Fig2}
\end{figure}

It is useful to examine the photon yield produced by interaction with a single nucleus $j$. Collecting the above expressions, we find
\begin{align}
\Gamma^{\rm coh}_j=\frac{3Z^2\alpha}{(v/c)^2\gamma^2}\;\frac{\kappa_r^2}{\omega_0\kappa}\;K_1^2\left(\frac{\omega_0 R_{jp}}{v\gamma}\right),
\label{Gj}
\end{align}
which we use to calculate the results shown in Fig.~\ref{Fig2}(a,b). We observe a smooth increase in the photon yield with increasing velocity, as well as a sharp drop as the beam moves away from the nucleus, in accordance with the small-distance ($R_{jp}\ll v\gamma/\omega_0$) behavior of Eq.~\eqref{Gj}, $\Gamma^{\rm coh}_j\propto1/R_{jp}^2$, independent of $v$. We note that BR emission can exceed the photon yield of the nuclear-excitation mechanism by several orders of magnitude (Fig.~\ref{Fig2}(a,b), dashed curves, see Appendix), although the strength of the BR mechanism scales as $M^{-2}$ with the mass of the probe $M$ and can therefore be neglected for massive ions. However, despite the higher number of photons generated via BR, the narrowness of the nuclear resonance renders a larger emission density within its spectral width [Fig.~\ref{Fig2}(c)]. In addition, BR emission takes place while the projectiles are traversing the material, and therefore, it can be neatly separated from $\gamma$-ray emission, which extends over a much longer period determined by the lifetime of the nuclear resonance [Fig.~\ref{Fig2}(d)], in the range of nanoseconds for the configuration under study. We therefore think of a bunched beam of electrons or ions, with a bunch duration in the sub-nanosecond range, as a way to excite the nuclei for subsequent reading through $\gamma$-ray emission over a longer time of several nanoseconds, using a fast detector to discriminate the arrival time of the $\gamma$ rays with respect to the exciting bunch.

Incoherent $\gamma$-ray emission can also take place as noted above, with an angular profile that should be equivalent to that generated by magnetic dipoles that are randomly oriented on a plane perpendicular to the local magnetic field (see Appendix). Considering this, as well as the probability of such incoherent processes relative to coherent emission, we find the probability of the former to have the angular distribution $\Gamma^{\rm incoh}(\Omega)=(3/16\pi)(1/f-1)\sum_j \left[1+\sin^2\theta\sin^2(\varphi-\varphi_{jp})\right]\,\Gamma^{\rm coh}_j$ with $\Gamma^{\rm coh}_j$ given by Eq.~\eqref{Gj}, which is therefore broad and featureless, so we dismiss it and concentrate instead on the sharp angular peaks associated with the nuclear SP coherent emission.

When the beam traverses a crystal lattice, the sum over nuclei can be analytically performed by separately computing it for each atomic plane. This is conveniently done by recasting the amplitude $\gb(\Omega)$ [Eq.~\eqref{gO}] in reciprocal space, so that we are left with a sum over the two-dimensional reciprocal lattice vectors $\Gb$, while the sum over layers produces the sharp angular profiles described by Eq.~\eqref{SP}. Following the procedure detailed in the Appendix for a film consisting of $N$ atomic planes with (100) surface orientation and body-centered cubic crystal symmetry, as appropriate for solid iron, we find an angle-resolved $\gamma$-ray SP emission probability
\begin{align}
\Gamma^{\rm coh}(\Omega)=\sum_n \Gamma^{\rm coh}_n(\varphi)\;\delta(\cos\theta-\cos\theta_n),
\nonumber
\end{align}
emerging along polar directions $\theta$ determined by Eq.~\eqref{SP} and with an azimuthal distribution given by
\begin{align}
&\Gamma^{\rm coh}_n(\varphi)\approx N\;\frac{18\pi^2Z^2\alpha\,\kappa_r^2}{ac(\omega_0a/c)^4\kappa} \label{Gm}\\
&\times{\sum_\Gb}' \; \frac{|\kb_\parallel+\Gb|^2\cos^2\theta_n+\big|(\kb_\parallel+\Gb)\cdot\rr\big|^2}{\left[|\kb_\parallel+\Gb|^2+(\omega_0/v\gamma)^2\right]^2}. 
\nonumber
\end{align}
Here, $a$ is the lattice constant, the sum $\sum'$ runs over two-dimensional reciprocal lattice vectors $\Gb=(2\pi/a)(i,j)$ subject to the condition that $i+j+n$ is an even integer number, and $\kparb$ is the projection of $\kb_0$ on the surface plane.

To obtain Eq.~\eqref{Gm}, we have averaged the emission probability over the impact parameter $\Rb_p$ under the assumption that the electron or ion beam extends over many crystal periods in the transverse direction. The resulting $\Gb$ sum exhibits a logarithmic divergence arising from large $G$ contributions and behaving as $\sim\int_{G<G_{\rm max}} d^2\Gb/G^2\sim\log G_{\rm max}$ when restricting the sum by a cutoff $G_{\rm max}$. In physical terms, this divergence is associated with close encounters between the probe and the nuclei, which are in fact avoided due to their Coulomb interaction. We thus express the cutoff $G_{\rm max}\sim1/R_{\rm min}$ in terms of the minimum impact parameter $R_{\rm min}$ with respect to the nuclear positions. Without entering into the details of the transverse modulation of the beam as it propagates along the crystal, we provide results obtained by computing Eq.~\eqref{Gm} for different values of $R_{\rm min}$, keeping in mind that this parameter is limited by the transverse beam energy, which is in turn $E_\perp\sim\theta_{\rm inc}^2E_0$, where $E_0$ is the longitudinal kinetic energy and we consider a small incidence angle $\theta_{\rm inc}$ relative to the rows of nuclei. For our Fe-57 target (atomic number $Z_{\rm Fe}=26$), the relevant transverse Coulomb energy is $E_{\rm Coul}=(2Z_{\rm Fe}e^2/a)\log(a/2R_{\rm min})$, corresponding to the difference in the potential created by a (100) atomic row at distances $R_{\rm min}$ and $a/2$ from it (i.e., the minimum and maximum separations from any such row inside the crystal) after averaging over positions along the row direction. In addition, we expect an angular broadening of the emission profile $\sim\theta_{\rm inc}$, so taking $\theta_{\rm inc}=2^\circ$ and $E_0=1\,$MeV as reasonable parameters, the condition $E_\perp\sim E_{\rm Coul}$ leads to $R_{\rm min}\sim1\,$pm, which should be considered as a plausible minimum distance, while even smaller values should be reachable with more energetic probes and similarly small beam inclinations or divergences.

\begin{figure}[htpb]
\centering
\includegraphics[width=0.45\textwidth]{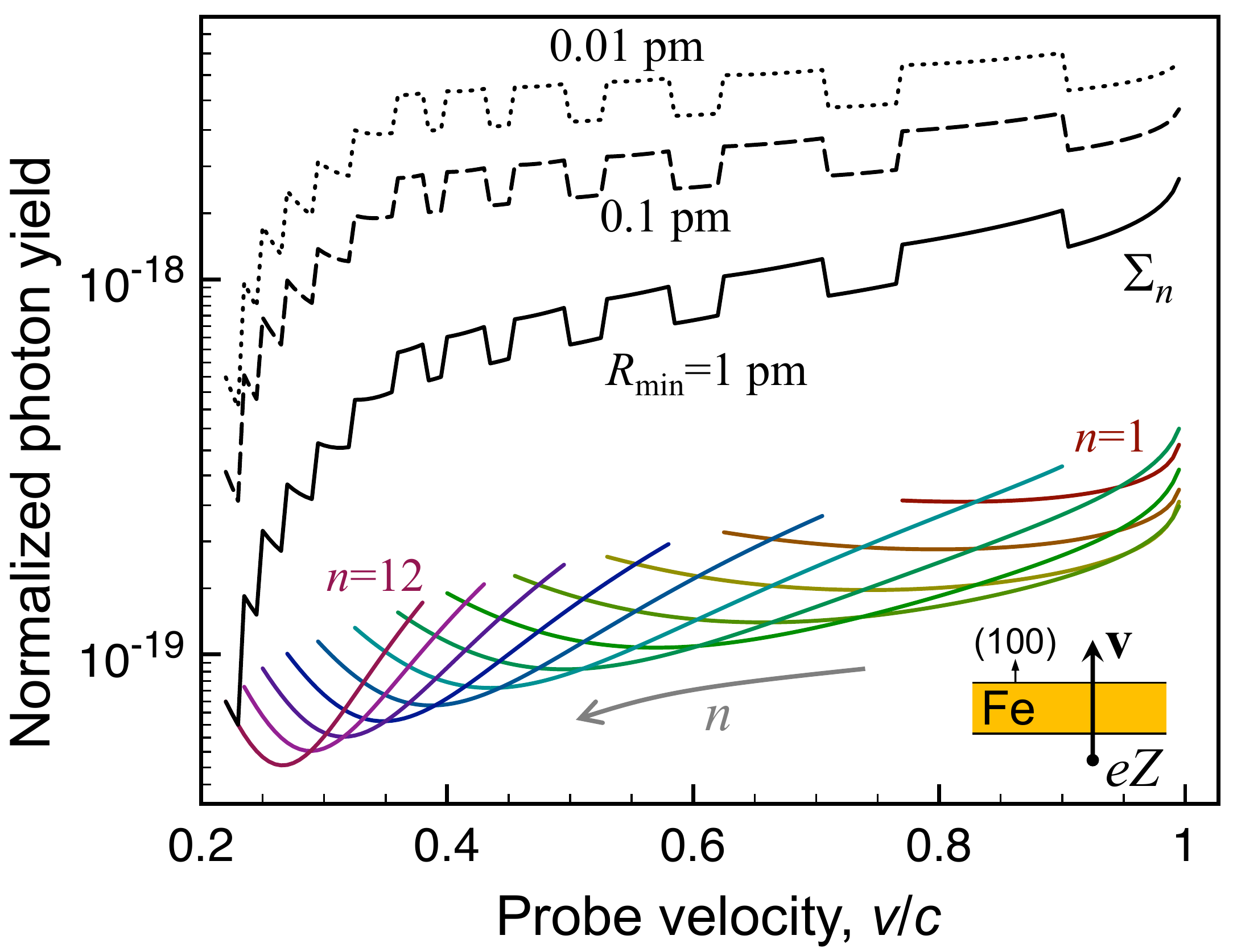}
\caption{\textbf{Coherent SP $\gamma$-ray emission.} We show the photon yield produced from a Fe-57 crystal film oriented perpendicularly to the (100) direction when it is traversed by normally-impinging electrons or ions as a function of probe velocity. The yield is divided by $Z^2$ and normalized per incident particle and per atomic layer in the film. We show results for three different values of the minimum beam-nucleus distance $R_{\rm min}$ summed for coherence orders $n=1-12$, as well as the decomposition of the $R_{\rm min}=1\,$pm yield in the contribution of different $n$'s (lower curves, see labels, integrated over azimuthal angles).}
\label{Fig3}
\end{figure}

The photon yield calculated from the angular integral of Eq.~\eqref{Gm} shows the expected logarithmic increase with decreasing $R_{\rm min}$, as well as an overall increase with particle velocity $v$, as we show in Fig.~\ref{Fig3}, where we plot the yield divided by $Z^2$ for particles of charge $eZ$ and normalized to the number of (100) atomic layers $N$ in the Fe-57 film. Interestingly, a rich structure of emission is found as a function of $v$, and in particular, different coherence orders $n$ produce commensurate contributions to the emission up to a large order (Fig.~\ref{Fig3}, lower curves), each of them associated with different emission angles according to Eq.~\eqref{SP}. The sum over $n$'s (black curves) exhibits sudden jumps as a function of $v$ that reflect the above-mentioned change in the parity of the reciprocal lattice vectors [see condition on $\Gb$ in Eq.~\eqref{Gm}] that contribute at consecutive orders. Incidentally, for the Fe-57 film under consideration, the lattice constant $a\approx2.86\,${\AA} is substantially larger than the photon wavelength $\lambda\approx0.86\,${\AA} and the cutoff parameter (e.g., $v\gamma/\omega_0\approx0.28${\AA} for $v=0.9\,c$), so we anticipate a weak azimuthal dependence in the angular distribution of the emission from Eq.~\eqref{Gm}, which we corroborate numerically. An azimuthal dependence is however encountered for ultrarelativistic particles ($v\lesssim c$, not shown). Under conditions previously explored in experiment \cite{KDD93}, we consider relavistic ions with $v$ close to $c$ and $Z\sim10$, which can travers a film thickness of several microns ($N\sim10^4$) and reach small impact parameters $R_{\rm min}$, so the photon yield per ion resulting by multiplying the result in Fig.~\ref{Fig3} by $N\,Z^2$ is $\sim10^{-11}$, which should be detectable considering the noted delay between the time of ion bunch bombardment and the emission of $\gamma$ photons.

\section{Conclusions}

In conclusion, a periodic array of nuclei in a solid crystal film can be coherently excited by electrons or ions traversing the material, giving rise to the emission of $\gamma$ rays along well-defined directions in analogy to the SP effect. Although we explore this effect here for Fe-57, it should also be observable in other nuclei, which configure a vast range of excitation energies and lifetimes \cite{IAEA}. The predicted yield of nuclear SP emission is predicted to increase with the mass of the probe, which should enable closer collisions with the nuclei in the structure. Further improvement of the yield could be obtained by operating under channeling conditions similar to what happens in the Okorokov effect \cite{OTK1973,KDD93,paper007}. Besides the fundamental interest lying in the demonstration of coherence between different nuclei in a sample by means of the observation of sharp angular emission profiles, our results hold potential as a characterization technique that should shed light into the spatial distribution of different isotopes in a material. The fact that $\gamma$ rays are released from the sample over an interval controlled by the lifetime of the nuclear excitation, extending well beyond the time of passage of the beam, suggests a strategy for storing energy in the form of nuclear excitations, which can be later liberated as $\gamma$-ray emission along narrowly peaked angular directions.

\appendix 
\section*{APPENDIX} 



In this appendix, we present detailed derivations of the coherent magnetic dipolar polarizability describing light scattering by Fe-57 and Dy-161 nuclei, the Smith-Purcell $\gamma$-ray emission from solid crystal targets containing these types of atoms, and the incoherent photon emission probability produced by electron or ion bombardment.

\maketitle 
\date{\today} 

\section{Coherent and incoherent light scattering by Fe-57 and Dy-161 nuclei}
\label{sec1}

The ground state of an Fe-57 nucleus has a total angular number $j_g=1/2$. We are interested in excitations to the state of energy $\hbar\omega_0=14.4129\,$keV in that nucleus, which has an total angular momentum $j_e=3/2$ and a measured lifetime $1/\kappa=142\,$ns \cite{IAEA} (for $1/e$ decay). Radiative coupling between the ground and excited states is mediated by magnetic-dipole (M1) and electric-quadrupole (E2) terms, with a ratio of their respective strengths given by E2/M1$\,\approx0.002$ \cite{IAEA}, so we ignore the E2 channel and assume pure magnetic-dipole transitions. In addition, the excited state can decay nonradiatively by transferring energy to electrons in the system (the so-called internal conversion mechanism) with a ratio of nonradiative-to-radiative rates $\alpha_{\rm IC}=8.544$ for the M1 channel \cite{KBT08,BrIcc}.

By analogy to electronic levels in atoms, we represent nuclear sublevels through a specific realization such as \cite{M1966}
\[|lj\mu\rangle=\sum_{s=-1/2}^{1/2}C_{lj\mu s}Y_{l\mu-s}|s\rangle,\]
describing the coupling between spin states $|s=\pm1/2\rangle$ and orbital states with the symmetry of the spherical harmonics $Y_{lm}$, characterized by angular momentum numbers $(l,m)$ (see below on the choice of $l$). Here, $j$ is the total angular momentum number of the nuclear level and $\mu=-j,\cdots,j$ labels the sublevels. These mixed states involve the nonvanishing half-integer Clebsch-Gordan coefficients
\begin{widetext}
\begin{align}
C_{lj\mu s}=\left\{\begin{array}{lll}
\quad\!\!\sqrt{(j+\mu+1)/2(j+1)}, &\quad\quad l=j+1/2, &\quad\quad s=-1/2, \\
-\sqrt{(j-\mu+1)/2(j+1)}, &\quad\quad l=j+1/2, &\quad\quad s=\;\;\,1/2, \\
\quad\!\!\sqrt{(j-\mu)/2j}, &\quad\quad l=j-1/2, &\quad\quad s=-1/2, \\
\quad\!\!\sqrt{(j+\mu)/2j}, &\quad\quad l=j-1/2, &\quad\quad s\;\;\,=1/2.
\end{array}\right. \nonumber
\end{align}
Dipolar coupling to a time-varying external magnetic field is realized through matrix elements sharing a common radial part and with angular components given by $\langle l'j'\mu'|\rr|lj\mu\rangle$. Expressing the radial unit vector as
\[\rr=\sqrt{4\pi/3}\left\{(1/\sqrt{2})\left[Y_{1,-1}(\Omega)-Y_{11}(\Omega)\right]\,\xx+(\ii/\sqrt{2})\left[Y_{1,-1}(\Omega)+Y_{11}(\Omega)\right]\,\yy+Y_{10}(\Omega)\,\zz\right\},\]
where $\Omega$ denotes the direction of $\rr$, we can readily obtain the matrix elements in terms of
\[\langle l'j'\mu'|Y_{1m}|lj\mu\rangle=\sum_s C_{l'j'\mu's}C_{lj\mu s}\int d\Omega \; Y^*_{l',\mu'-s}(\Omega)Y_{1m}(\Omega)Y_{l,\mu-s}(\Omega),\]
which vanishes unless $l'-j'=l-j$. The result is independent of the actual realization (i.e., the choice of $l=j\pm1/2$), and in particular, for the Fe-57 system ($j'=j_e=3/2$ and $j=j_g=1/2$), using the notation $|e_{\mu'}\rangle=|1,3/2,\mu'\rangle$ and $|g_\mu\rangle=|0,1/2,\mu\rangle$ for the excited and ground sublevels, respectively, we find
\begin{align}
\langle e_{\mu'}|Y_{lm}|g_\mu\rangle=\delta_{\mu',\mu+m}\frac{1}{\sqrt{4\pi}}\times\left\{\begin{array}{cll}
1 &\quad\quad \mu'=\pm3/2, &\quad\quad m\neq0, \\
\sqrt{2/3}, &\quad\quad \mu'=\pm1/2, &\quad\quad m=0, \\
\sqrt{1/3}, &\quad\quad \mu'=\pm1/2, &\quad\quad m\neq0, \\
0, &\quad\quad \text{elsewhere.} &
\end{array}\right. \nonumber
\end{align}
\end{widetext}
Squaring these matrix elements, the relative strengths of the transitions connecting different sublevels are found to be as indicated by the orange numbers in the following diagram (reproduced from Fig.~\ref{Fig1}(d) in the main text):\\
\begin{center} \includegraphics[width=0.45\textwidth]{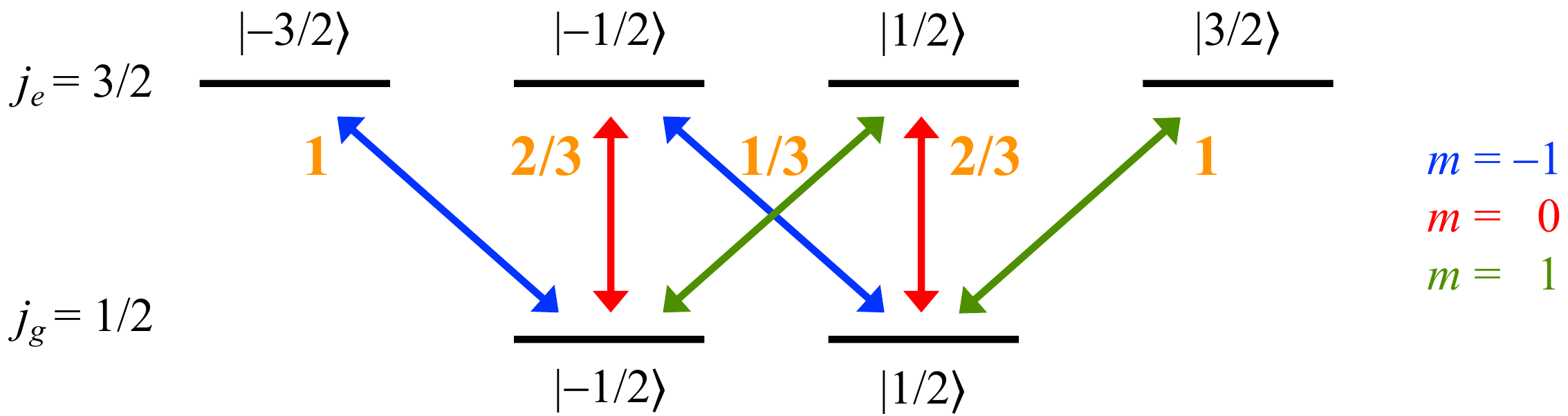} \\ \end{center}
Blue, red, and green double arrows correspond to transitions in which $m=\mu'-\mu=-1$, 0, and 1, respectively. The sum of strengths for downward transitions from any given excited state to the accessible ground states is 1, whereas the sum of upward arrows from any of the two ground states is 2. These numbers are in the expected ratio $(2j_e+1)/(2j_g+1)=2$, reflecting the fact that more final states are available for excitation than for de-excitation.

\begin{table*}[t]
\caption{Parameters entering the magnetic dipolar polarizability of Fe-57 and Dy-161 in Eq.\ (\ref{alphaM}). Here, $\omega_0$, $\kappa$, $\alpha_{\rm IC}$, $j_g$, and $j_e$ are taken from Refs.\ \onlinecite{IAEA,KBT08,BrIcc}, while $f$ and $\kappa_r$ are derived as explained in this section.}
\begin{tabular}{l|ccccccc}
       &$\hbar\omega_0$ (keV) &$1/\kappa$ (s) &$1/\kappa_r$ (s)    &$\alpha_{\rm IC}$ &$f$ &$j_g$ &$j_e$\\ \hline
Fe-57  &14.4129      &$1.42\times10^{-7}$     &$2.03\times10^{-6}$ &8.544             &2/3 &1/2   &3/2  \\
Dy-161 &43.8201      &$1.20\times10^{-9}$     &$3.17\times10^{-8}$ &4.213             &4/9 &5/2   &7/2  \\
\end{tabular}
\label{Table1}
\end{table*}

In accordance with the discussion presented above, we describe the coherent scattering of electromagnetic fields by the nucleus in terms of a magnetic dipolar polarizability. Then, a given Fe-57 nucleus prepared in either of the two ground-state sublevels ($|g_{-1/2}\rangle$ or $|g_{1/2}\rangle$) is characterized by an anisotropic polarizability in which each $m$ component takes a different value propotional to the corresponding transition strength in the above diagram. However, by averaging over the two possible values of $\mu=\pm1/2$ in the ground-state sublevels, the magnetic polarizability becomes isotropic and given by
\begin{align}
\alpha_M(\omega)\approx\frac{3}{4k^3}\frac{\kappa_r}{\omega_0-\omega-\ii\kappa/2}, \label{alphaM}
\end{align}
where $\kappa_r$ is the {\it coherent} radiative decay rate and $k=\omega_0/c$. The internal conversion mechanism reduces the radiative transition component of the decay rate by a factor $1+\alpha_{\rm IC}$ relative to $\kappa$ (i.e., nonradiative decay channels contribute to broaden the resonance with a rate $\kappa$ deduced from the measured decay time, but only a fraction $1/(1+\alpha_{\rm IC})$ of all decay channels is associated with the emission of radiation). In addition, just an average fraction $f=2/3$ of those radiative decay channels [the $m$-independent average of the strengths (orange numbers) associated with each $m$ in the above diagram] is coherent in the sense that they leave the system in the original initial state, whereas the remaining $1-f$ fraction corresponds to incoherent emission of photons accompanied by a change in $\mu$. Combining these factors, we set $\kappa_r=\kappa\times f/(1+\alpha_{\rm IC})\approx1/(2.03\,\mu\text{s})$ for Fe-57. The parameters entering or affecting Eq.\ (\ref{alphaM}) are summarized in Table\ \ref{Table1}.

As a consequence of the above diagram, we find that, upon excitation by an optical magnetic field directed along a given direction $\nn$, the incoherent radiative decay channels lead to photon emission with an angular profile corresponding to the average of the emission from magnetic dipoles oriented perpendicularly to $\nn$. For example, when exciting with $m=0$ (i.e., $\nn=\zz$) from an initial sublevel $|\pm1/2\rangle$, incoherent radiative decay to the other sublevel $|\mp1/2\rangle$ involves emission of photons with $m=\pm1$, whose average is also obtained as the one over the emission of dipoles oriented along $\xx$ and $\yy$ (i.e., in the plane perdendicular to the exciting magnetic field). Because the averaged nuclear response is isotropic, this argument applies to any orientation $\nn$ of the applied field. We use this result below to calculate the incoherent emission probability from the coherent one in a straightforward fashion.

For Dy-161 nuclei, the excited state of energy $\hbar\omega_0=43.8201\,$keV is also dominated by magnetic dipole coupling, with a ratio E2/M1$\,\sim10^{-9}$ and a higher degree of degeneracy in the ground and excited states ($j_g=5/2$ and $j_e=7/2$). An analysis similar to the one carried out above for iron leads to the following diagram for Dy-161: \\
\begin{center} \includegraphics[width=0.45\textwidth]{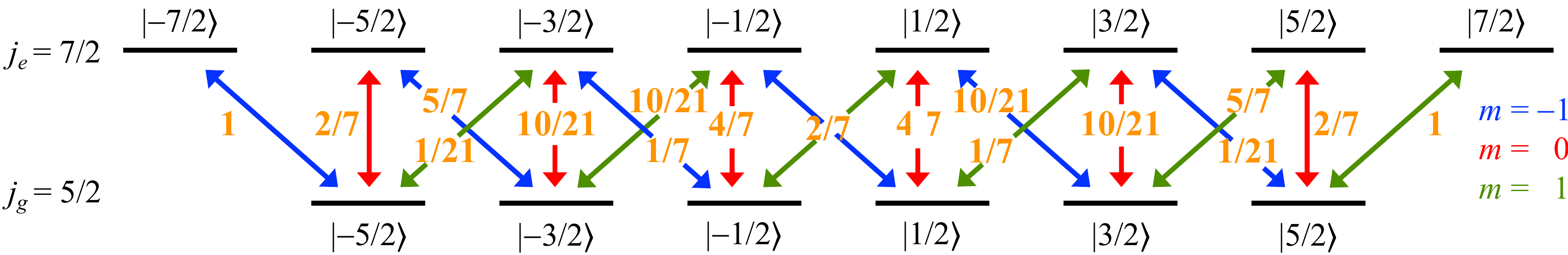} \\ \end{center}
From these transition strengths, we find a coherent fraction $f=4/9$, which combined with tabulated values of $\kappa$ and $\alpha_{\rm IC}$ taken from Refs.\ \onlinecite{IAEA,KBT08,BrIcc}, leads to the parameter list given in Table\ \ref{Table1}, although in this case $\kappa_r$ needs to be reduced by a factor of 2.25 to account for decay mediated by transitions to an intermediate state that do not contributed to the studied coherent radiative channel. Likewise, we can describe coherent light scattering by Dy-161 nuclei through the magnetic polarizability in Eq.\ (\ref{alphaM}).

\section{Coherent $\gamma$-ray cathodoluminescence}
\label{sec2}

We work in the nonrecoil approximation (i.e., the electron or ion velocity $\vb$ is assumed to remain constant during the interaction with the target), which is valid when the excitation energies are small compared with the kinetic energy of the probe, so that it is not strongly deflected by close collisions with the nuclei. In addition, the probability that a given electron or ion emits one $\gamma$-ray photon is much smaller than unity, so we can describe the interaction within first-order perturbation theory. Moreover, we consider monochromatic, collimated beams, moving along a well-defined direction. Under these conditions, a rigorous quantum-mechanical treatment of the probe-target interaction problem \cite{paper371} shows that the excitation probability is identical with the one calculated by assimilating the electron or ion to the classical evanescent electromagnetic field that accompanies a moving point charge $eZ$ (e.g., $Z=-1$ for electrons). For beams of finite extension along the transverse directions ($\perp\vb$), this probability needs to be averaged over the beam density profile. In the present work, where the studied nuclei respond through their magnetic polarizability, we just need to plug the external magnetic field. Considering a trajectory $\rb=\Rb_p+\vb t$, where $\Rb_p\perp\vb$ acts as an impact parameter, the probe produces an evanescent magnetic field $\Hb^{\rm ext}(\rb,t)=(2\pi)^{-1}\int d\omega\;\ee^{-\ii\omega t}\Hb^{\rm ext}(\rb,\omega)$ with spectral components \cite{paper149}
\begin{align}
\Hb^{\rm ext}(\rb,\omega)=\frac{2eZ\omega}{vc\gamma}\,K_1\left(\frac{\omega |\Rb-\Rb_p|}{v\gamma}\right)\ee^{\ii\omega z/v}\,
\hat\varphi,
\label{Hext}
\end{align}
where we use cylindrical coordinates $\rb=(R,\varphi,z)$ and transverse coordinates $\Rb=(x,y)$, and $\gamma=1/\sqrt{1-v^2/c^2}$ is the relativistic Lorentz factor.

For a collection of nuclei distributed at positions $\rb_j$, coherent $\gamma$-ray emission is mediated by the frequency-dependent magnetic dipoles $\mb_j(\omega)=\alpha_M(\omega)\Hb^{\rm ext}(\rb_j,\omega)$ created in response to the field in Eq.\ (\ref{Hext}) via the polarizability $\alpha_M(\omega)$ given in Eq.\ (\ref{alphaM}). The dipole induced at a nucleus $j$ produces an electric field $-k^2(\rr\times\mb_j)\,\ee^{\ii k|\rb-\rb_j|}/|\rb-\rb_j|$. In the far-field limit ($kr\gg1$ and $r\gg r_j$), this expression can be approximated as $-k^2(\rr\times\mb_j)\,\ee^{-\ii\kb\cdot\rb_j}\,\ee^{\ii kr}/r$ with $\kb=k\rr$. Summing over all dipoles, the resulting induced (emitted) electric field reduces to
\begin{align}
\Eb^{\rm ind}(\rb,\omega)=\fb(\omega) \frac{\ee^{\ii kr}}{r},
\nonumber
\end{align}
where
\begin{align}
\fb(\Omega,\omega)=&-\alpha_M(\omega)\frac{2eZk^3}{v\gamma} \label{fWw}\\
&\times\sum_j K_1\left(\frac{\omega R_{jp}}{v\gamma}\right)\,\ee^{\ii\omega z_j/v}\,\ee^{-\ii\kb\cdot\rb_j}\,\rr\times\hat\varphi_{jp}
\nonumber
\end{align}
is the emission amplitude. Here, $\Omega$ denotes the direction of $\rr$ and we have defined the relative transverse coordinates $\Rb_{jp}=\Rb_j-\Rb_p$ and the azimuthal vectors $\hat\varphi_{jp}\perp\Rb_{jp},\,\zz$. Following the procedure described in Ref.\ \onlinecite{paper149}, the coherent photon emission probability $\Gamma^{\rm coh}$ can be obtained by integrating the radial Poynting vector over time and directions of emission $\Omega$, and then dividing the result by the photon energy $\hbar\omega$. We find $\Gamma^{\rm coh}=\int d\Omega \,\Gamma^{\rm coh}(\Omega)$ with
\begin{align}
\Gamma^{\rm coh}(\Omega)=\frac{c}{4\pi^2\hbar}\int_0^\infty \frac{d\omega}{\omega} \left|\fb(\Omega,\omega)\right|^2.
\label{GW1}
\end{align}
Inserting Eq.\ (\ref{fWw}) into Eq.\ (\ref{GW1}), we can carry out the $\omega$ integral by approximating $\omega\approx\omega_0$ outside $\alpha_M(\omega)$ because the emission concentrates around a narrow spectral range of width $\kappa\ll\omega_0$ around that frequency. This leads to
\begin{align}
\Gamma^{\rm coh}(\Omega)=\frac{9Z^2\,\alpha}{8\pi(v/c)^2\gamma^2}\;\frac{\kappa_r^2}{\omega_0\kappa}\;\left|\rr\times\gb(\Omega)\right|^2,
\label{GcohW}
\end{align}
where $\alpha\approx1/137$ is the fine structure constant and
\begin{align}
\gb(\Omega)=\sum_j K_1\left(\frac{\omega_0 R_{jp}}{v\gamma}\right)\,\ee^{\ii\omega_0z_j/v}\,\ee^{-\ii\kb_0\cdot\rb_j}\,\hat\varphi_{jp}
\label{gW}
\end{align}
depends on the emission direction $\Omega$ through $\kb_0=(\omega_0/c)\,\rr$.

For a single nucleus $j$, the angle-integrated probability reduces to
\begin{align}
\Gamma^{\rm coh}_j=\frac{3Z^2\alpha}{(v/c)^2\gamma^2}\;\frac{\kappa_r^2}{\omega_0\kappa}\;K_1^2\left(\frac{\omega_0 R_{jp}}{v\gamma}\right),
\label{Gj}
\end{align}
which decays exponentially at large nucleus-beam distances $R_{jp}$ but diverges as $\sim1/R_{jp}$ for close encounters.

\section{Smith-Purcell $\gamma$-ray emission from a solid crystal target}
\label{sec3}

The coherent emission from nuclei arranged in a crystal lattice gives rise to $\gamma$-ray emission along directions determined by the condition of constructive far-field interference, similar to the Smith-Purcell effect for grating structures \cite{SP1953,V1973a}. We consider a homogeneous crystalline film consisting of $N$ atomic planes with one emitting nucleus in each two-dimensional (2D) unit cell of each of those planes. For simplicity, we take the film to be normal to the $z$ axis (i.e., we consider normally impinging beams). The crystal structure can then be defined in terms of the 2D reciprocal lattice and the three-dimensional displacement vector $\bb=\bb_\parallel+b_z\zz$ separating the reference nuclei in two consecutive atomic planes.

At this point, we need to multiplex the sum over nuclei sites $\rb_j$ in Eq.\ (\ref{gW}) into two sums: one over 2D real-lattice sites $\Rb_j$ and the other over atomic planes $z=z_l$, keeping in mind that we replace $\rb_j$ by $\Rb_j+l\bb$. In addition, the 2D lattice sum can be transformed into a sum over 2D reciprocal lattice vectors $\Gb$ by consecutively applying the identities
\begin{align}
K_1(\Delta R)\hat\varphi &= \frac{1}{\Delta}(\xx\partial_y-\yy\partial_x)K_0(\Delta R), \nonumber\\
K_0(\Delta R) &= \frac{1}{2\pi}\int d^2\Qb\frac{\ee^{\ii\Qb\cdot\Rb}}{Q^2+\Delta^2}, \nonumber\\
\sum_{\Rb_j} \ee^{\ii\Qb\cdot\Rb_j} &= \frac{(2\pi)^2}{A}\sum_\Gb \delta(\Qb-\Gb), \nonumber
\end{align}
where $A$ is the unit cell area and $\Delta=\omega_0/v\gamma$. We find
\begin{align}
\gb(\Omega)=-\frac{2\pi\ii v\gamma}{A\omega_0} \,\sum_\Gb&\ee^{-\ii(\kb_\parallel+\Gb)\cdot\Rb_p}\,\frac{|\kb_\parallel+\Gb|}{|\kb_\parallel+\Gb|^2+(\omega_0/v\gamma)^2} \nonumber\\
&\times \hat\varphi_{\kb_\parallel+\Gb} \; S_\Gb,
\label{gWG}
\end{align}
where $\kb_\parallel$ is the in-plane component of $\kb_0=(\omega_0/c)\rr$, we define the azimuthal vector $\hat\varphi_\Qb=(-Q_y\xx+Q_x\yy)/Q$, and the coefficient
\begin{align}
S_\Gb=\sum_\ell\exp\left\{\ii\ell\left[(\omega_0b_z/c)(c/v-\cos\theta)+\Gb\cdot\bb_\parallel\right]\right\}
\label{SG}
\end{align}
encapsulates the sum over atomic planes and depends on the polar angle of emission $\theta$ relative to $\zz$.

We now insert Eq.\ (\ref{gWG}) into Eq.\ (\ref{GcohW}) and take the average of the result over the beam impact parameter $\Rb_e$ (i.e., we average over lateral film positions). This allows to simplify the double sum over reciprocal lattice vectors emerging when squaring $\gb(\Omega)$ in Eq.\ (\ref{GcohW}) by applying $A^{-1}\int_A d^2\Rb_p\,\left|\sum_\Gb\ee^{-\ii\Gb\cdot\Rb_p}\ab_\Gb\right|^2=\sum_\Gb|\ab_\Gb|^2$, which is a consequence of the identity $A^{-1}\int_A d^2\Rb_p \ee^{\ii(\Gb-\Gb')\cdot\Rb_p}=\delta_{\Gb,\Gb'}$, where the integrals extend over a 2D unit cell. In addition, the geometric sum over atomic planes in Eq.\ (\ref{SG}) can also be carry out analytically to yield $|S_\Gb|^2=\sin^2(N\xi_\Gb/2)/\sin^2(\xi_\Gb/2)$, where \[\xi_\Gb=(\omega_0b_z/c)(c/v-\cos\theta)+\Gb\cdot\bb_\parallel\] and $N$ is the number of atomic layers. For $N\gg1$, we have $|S_\Gb|^2\approx2\pi N\sum_m\delta(\xi_\Gb-2\pi m)$, where $m$ runs over integer numbers. Putting these elements together, we obtain
\begin{align}
\Gamma^{\rm coh}(\Omega)\approx& N\;\frac{9\pi^2Z^2\alpha\, c^2\kappa_r^2}{A^2\omega_0^3\kappa}\; \sum_\Gb \frac{|\kb_\parallel+\Gb|^2}{\left[|\kb_\parallel+\Gb|^2+(\omega_0/v\gamma)^2\right]^2} \nonumber\\
&\times\left|\rr\times\hat\varphi_{\kb_\parallel+\Gb}\right|^2\; \sum_m\delta\left(\xi_\Gb-2\pi m\right). \nonumber
\end{align}
The $\delta$-function in each $m$ term determines a polar cone of emission through the condition $\xi_\Gb=2\pi m$, where the dependence on $\Gb$ can be partially or completely eliminated by redefining $m$ for different types of crystal lattices. In particular, for a (100) film made of a simple cubic crystal with lattice constant $a$, we can take $\bb_\parallel=0$ and $b_z=a$, so the $\Gb$ dependence disappears from $\xi_\Gb$ and we recover the familiar Smith-Purcell condition $\cos\theta_n=c/v-n\lambda/a$ in terms of the light wavelength $\lambda$ and the interatomic plane spacing $a$.

Iron forms a bcc lattice, so considering a (100) film orientation, we can take $\bb_\parallel=(a/2,a/2)$, $b_z=a/2$, $A=a^2$, and $\Gb=(2\pi/a)(i,j)$, where $a=2.856\,{\AA}$ is the lattice constant, while $i$ and $j$ run over integer numbers. For such Fe(100) film, we have $\Gb\cdot\bb_\parallel=(i+j)\,\pi$, so we can redefine $n=2m-i-j$ and separate the emission as a sum over components directed along the cones determined by
\begin{align}
\cos\theta_n=c/v-n\lambda/d
\label{costhm}
\end{align}
according to
\begin{align}
\Gamma^{\rm coh}(\Omega)=\sum_n \Gamma^{\rm coh}_n(\varphi)\;\delta(\cos\theta-\cos\theta_n),
\nonumber
\end{align}
where $d=a$ gives the periodicity along the out-of-plane direction, $\lambda=2\pi c/\omega_0$ is the emission wavelength, and
\begin{align}
\Gamma^{\rm coh}_n(\varphi)\approx &N\;\frac{18\pi^2Z^2\alpha\,\kappa_r^2}{dc(\omega_0a/c)^4\kappa}\; {\sum_\Gb}' \frac{|\kb_\parallel+\Gb|^2}{\left[|\kb_\parallel+\Gb|^2+(\omega_0/v\gamma)^2\right]^2} \nonumber\\
&\times\left|\rr\times\hat\varphi_{\kb_\parallel+\Gb}\right|^2 \label{Gm}
\end{align}
is the probability density as a function of the azimuthal angle of emission $\varphi$. An implicit dependence on $\theta_n$ arises in Eq.\ (\ref{Gm}) through $\kparb=k\sin\theta_n\;(\cos\varphi,\sin\varphi)$. Here, the prime in ${\sum}'$ indicates that the sum is restricted to $\Gb=(2\pi/a)(i,j)$ vectors such that $i+j+n$ ($=2m$) is an even integer.

For Dy-161, we consider a (100) film of DyN, which forms an fcc crystal structure with a Dy-Dy nearest-neighbors distance $a'=3.6\,{\AA}$ \cite{SJK15}, so we can readily apply the results in Eqs.\ (\ref{costhm}) and (\ref{Gm}) by defining 2D atomic planes with a square lattice of period $a=a'$ and vertical spacing $b_z=a'/\sqrt{2}$ (i.e., $d=a\sqrt{2}$).

\section{Incoherent $\gamma$-ray emission}
\label{sec4}

Each beam-excited nucleus can decay to a state that differs from the original ground state (i.e., to a different sublevel $|\mu\rangle$) by emitting one $\gamma$-ray photon. The probability $\Gamma^{\rm incoh}$ associated with this incoherent process is the sum of the probabilities contributed by all nuclei in the target. Then, as we argue in Sec.\ \ref{sec1}, each nucleus emits incoherently with an angular profile given by the average of magnetic dipoles oriented perpendicularly to the external magnetic field. For a given nucleus, since the magnetic field of the probe is parallel to the azimuthal direction $\hat\varphi_{jp}$ [see Eqs.\ (\ref{Hext}) and (\ref{fWw})], we have to average the field intensities produced by two magnetic dipoles of equal magnitude oriented along $\Rb_{jp}$ and $\zz$. This results in an angular profile $\propto|\rr\times\RR_{jp}|^2+|\rr\times\zz|^2=1+\sin^2\theta\sin^2(\varphi-\varphi_{jp})$, where the angles $\Omega=(\theta,\varphi)$ define the emission direction. Making use of the fact that the respective fractions of coherent and incoherent emission are $f$ and $1-f$ (see Sec.\ \ref{sec1} and Table\ \ref{Table1}), we can readily write the angle-resolved incoherent emission probability as
\begin{align}
\Gamma^{\rm incoh}(\Omega)=&\frac{9Z^2\alpha}{16\pi(v/c)^2\gamma^2}\;\frac{\kappa_r^2}{\omega_0\kappa}\;\left(\frac{1}{f}-1\right) \nonumber\\
&\times\sum_j K_1^2\left(\frac{\omega_0 R_{jp}}{v\gamma}\right)\;\left[1+\sin^2\theta\sin^2(\varphi-\varphi_{jp})\right].
\nonumber
\end{align}
To obtain this result, we have added the coherent emission probability coming from each nuclei $j$ [Eq.\ (\ref{Gj})], previously multiplied by both $(1-f)/f$ and the normalized angular profile.

\section{Bremsstrahlung emission}

Under the conditions here investigated, particles in the beam can be deflected by the Coulomb potential of the target nuclei, leading to Bremsstrahlung (BR) emission. The radiated energy can be expressed as $\int d\Omega\int_0^\infty d\omega\,\hbar\omega\,\Gamma^\mathrm{BR}(\Omega,\omega)$, where $\Gamma^\mathrm{BR}(\Omega,\omega)$ gives the spectral and angular distribution of the emission probability. For an arbitrary time-dependent velocity vector $\vb(t)$ and trajectory $\rb_p(t)$ of the particle, this probability admits the expression \cite{J99}
\begin{align}
\Gamma^\mathrm{BR}&(\Omega,\omega)=\frac{\alpha Z^2}{4\pi^2\omega} \label{BR}\\
&\times\left\vert\int_{-\infty}^\infty dt\;\ee^{\ii\omega\left[t-\rr\cdot\rb_p(t)/c\right]}\;\frac{d}{dt}\left[\frac{\rr\times\vb(t)/c}{1-\rr\cdot\vb(t)/c}\right]\right\vert^2.
\nonumber
\end{align}
Assuming that the interaction with a target nucleus only produces a small perturbation in the particle velocity relative to its initial value $\vb(-\infty)=v_0\zz$, we can retain corrections to first order in the Coulomb potential and write the equation of motion $\dot{\vb}(t)\approx\big(ZZ_ne^2/M\gamma\big)\big(\Rb+\zz\,v_0t/\gamma^2\big)/(R^2+v_0^2t^2)^{3/2}$, where $\Rb$ is the beam-nucleus separation vector, $eZ$ and $eZ_n$ are the probe and nucleus charges, $M$ is the mass of the projectile, the Lorentz factor $\gamma$ is evaluated from $v_0$, and we neglect any effect arising from electrons surrounding the nucleus in the actual material. We now expand Eq.~\eqref{BR} to first order in $\dot{\vb}$ and use the above equation of motion to work out the time integral, which yields the result
\begin{align}
\Gamma^\mathrm{BR}&(\Omega,\omega)=\frac{\alpha^3Z^4Z_n^2\hbar^2\omega}{\pi^2M^2\gamma^2v_0^4} \label{BR2}\\
&\times\left\vert\left(1-\beta\cos\theta\right)\rr\times\Fb+\beta\left(\rr\times\zz\right)\left(\rr\cdot\Fb\right)\right\vert^2,
\nonumber
\end{align}
where $\Fb=K_1\left(\zeta\right)\RR+(\ii/\gamma^2)K_0\left(\zeta\right)\zz$, $\zeta=(1-\beta\cos\theta)\,\omega R/v_0$, and $\beta=v_0/c$. We use Eq.~\eqref{BR2} to obtain the BR curves in Fig.~2 of the main text, as it gives reasonable results for the velocities under consideration (e.g., as compared to a virtual-quanta analysis \cite{J99}).


\section*{ACKNOWLEDGMENTS} 
This work has been supported in part by the European Commission (Horizon 2020 Grant No. 964591-SMART-electron), the European Research Council (Advanced Grant 789104-eNANO), the Spanish MICINN (PID2020-112625GB-I00 and SEV2015-0522), the Catalan CERCA Program, and Fundaci\'os Cellex and Mir-Puig. S.G acknowledges support from Google Inc.



\end{document}